\documentclass[manuscript]{acmart}
\usepackage{makecell}
\usepackage{multirow}
\usepackage{pdflscape}
\usepackage{longtable}
\usepackage{geometry}
\usepackage{graphicx}
\usepackage{subcaption}
\usepackage{caption}
\usepackage{booktabs}
\usepackage{array}
\usepackage{wrapfig}
\usepackage{numname}

\usepackage{microtype}
\usepackage{hyperref}
\usepackage{url}
\usepackage{booktabs}
\usepackage{makecell}
\usepackage{multirow}
\usepackage{pdflscape}
\usepackage{longtable}
\usepackage{graphicx}
\usepackage{subcaption}
\usepackage{caption}
\usepackage{booktabs}
\usepackage{array}
\usepackage{wrapfig}
\usepackage{amsmath}
\usepackage{enumitem}
\usepackage{setspace}
\usepackage{adjustbox}
\usepackage{lineno}
\usepackage{ulem}

\AtBeginDocument{%
  }

\setcopyright{acmlicensed}
\copyrightyear{2018}
\acmYear{2018}
\acmDOI{XXXXXXX.XXXXXXX}
\acmConference[Conference acronym 'XX]{Make sure to enter the correct
  conference title from your rights confirmation email}{June 03--05,
  2018}{Woodstock, NY}
\acmISBN{978-1-4503-XXXX-X/2018/06}

\begin{document}

\title{Free Lunch for User Experience: Crowdsourcing Agents for Scalable User Studies}

\author{Siyang Liu}
\email{lsiyang@umich.edu}
\orcid{0000-0001-6780-9419}
\affiliation{%
  \institution{Language and Information Technology Group, University of Michigan}
  \city{Ann Arbor}
  \state{Michigan}
  \country{USA}
}

\author{Sahand Sabour}
\affiliation{%
  \institution{Tsinghua University}
  \city{Beijing}
  \country{China}}

\author{Xiaoyang Wang}
\affiliation{%
  \institution{America Tencent}
  \city{Bellevue}
  \country{USA}
}

\author{Rada Mihalcea}
\affiliation{%
  \institution{Language and Information Technology Group, University of Michigan}
  \city{Ann Arbor}
  \state{Michigan}
  \country{USA}
  }
\email{mihalcea@umich.edu}

\renewcommand{\shortauthors}{Trovato et al.}

\begin{abstract}
User studies are central to user experience research, yet recruiting participant is expensive, slow, and limited in diversity. 
Recent work has explored using Large Language Models as simulated users, but doubts about fidelity have hindered practical adoption. 
We deepen this line of research by asking whether scale itself can enable useful simulation, even if not perfectly accurate.
We introduce Crowdsourcing Simulated User Agents, a method that recruits generative agents from billion-scale profile assets to act as study participants. 
Unlike handcrafted simulations, agents are treated as recruitable, screenable, and engageable across UX research stages.
To ground this method, we demonstrate a game prototyping study with hundreds of simulated players, comparing their insights against a 10-participant local user study and a 20-participant crowdsourcing study with humans. 
We find a clear scaling effect: as the number of simulated user agents increases, coverage of human findings rises smoothly and plateaus around 90\%.
12.8 simulated agents are as useful as one locally recruited human, and 3.2 agents are as useful as one crowdsourced human.
Results show that while individual agents are imperfect, aggregated simulations produce representative and actionable insights comparable to real users.
Professional designers further rated these insights as balancing fidelity, cost, time efficiency, and usefulness.
Finally, we release an agent crowdsourcing toolkit with a modular open-source pipeline and a curated pool of profiles synced from ongoing simulation research, to lower the barrier for researchers to adopt simulated participants. 
Together, this work contributes a validated method and reusable toolkit that expand the options for conducting scalable and practical UX studies.


\end{abstract}


\keywords{Do, Not, Use, This, Code, Put, the, Correct, Terms, for,
  Your, Paper}

\received{20 February 2007}
\received[revised]{12 March 2009}
\received[accepted]{5 June 2009}

\maketitle

\section{Introduction}
User studies are central to UX research, yet participant recruitment has long faced a trade-off: lab participants provide high-quality insights but are scarce and costly, while crowdsourced participants are faster to recruit but uneven in quality and still expensive at scale~\cite{spool2001testing, kittur2008crowdsourcing, 10.1145/2858036.2858498}. These constraints have limited how quickly researchers can evaluate and iterate on new technologies.

Recent work has begun exploring Large Language Models (LLMs) as simulated participants, offering the possibility of faster and cheaper studies~\cite{park2023generative, park2022social, xu2024ai, jin2025teachtune, lu2024generative, markel2023gpteach}. However, critiques of simulation highlight issues such as bias, limited diversity, value–action gaps, and low robustness~\cite{mihalcea2025ai, natalie2025not, rime2024interviewing, harding2024ai, kwok2024evaluating}. More importantly, some critiques imply that if simulations cannot accurately capture differences across individuals and groups, they should not be used~\cite{harding2024ai, kwok2024evaluating}. Concerns over simulation accuracy have thus hindered practical adoption.

We deepen this line of research by shifting the question: can scaling up simulations enable useful simulation, even if not perfectly accurate? Instead of focusing on the imperfections of individual agents, we ask whether aggregated outputs from large numbers of simulated participants can yield insights comparable to those from human studies.
To explore this, we introduce Crowdsourcing Simulated User Agents (CSUA), a method that extends the logic of crowdsourced studies from human workers to large-scale LLM-based agents. CSUA provides a modular pipeline—onboarding, screening, experiencing, and feedback—that treats agents as recruitable, screenable, and engageable across standard UX research stages. 
The pipeline builds on billion-scale profile assets to lower the barrier for adoption and reuse.
We validate this approach in a game prototyping study, constructing 240 simulated players from 2,900 candidate profiles and comparing their insights against two human baselines: a 10-participant local study and a 20-participant crowdsourced study. Results reveal a clear scaling effect: as the number of simulated participants increases, coverage of human-derived findings rises smoothly and plateaus around 90\%. We estimate that 12.8 simulated agents are as useful as one local participant, and 3.2 as useful as one crowdsourced participant. 
Professional designers further rated the simulated outputs as balancing fidelity, cost, time efficiency, and usefulness.

Our contributions are threefold:

\begin{enumerate}
    \item Method. We introduce a framework for crowdsourcing simulated participants, positioning them as study participants that can be recruited, screened, and engaged across UX research stages.
    \item Toolkit. We release an open-source, modular pipeline and curated profile pools to help researchers adopt and reuse this approach.
    \item Empirical insight. Through a game prototyping study, we show that while individual agents are imperfect, aggregated simulations yield representative and actionable insights comparable to human participants, validating scalability as a key driver of utility.
\end{enumerate}
We position simulated participants not as replacements for human studies, but as a complementary tool in the UX research toolkit—especially valuable in early-stage prototyping where speed, scale, and diversity matter most.

\section{Related Work}

\subsection{Challenges in Collecting User Experience Data}
User experience provides valuable insights into product design and development, but collecting it often comes with high financial and logistical costs due to reliance on human participants~\cite{spool2001testing, kittur2008crowdsourcing}.
Local (in-person) recruitment has long been a primary approach, as it offers strong control over process quality. 
However, scaling is limited: recruiting specific groups not locally available is difficult and costly, restricting both study size and participant diversity~\cite{barkhuus2007mice, hankerson2016does, madampe2025addressing}.
For instance, the median sample size reported in user studies published in the ACM CHI proceedings is only 12~\cite{10.1145/2858036.2858498}.
Crowdsourcing platforms such as MTurk and Prolific have emerged as complementary solutions to broaden recruitment and reduce costs~\cite{kittur2008crowdsourcing, 10.1371/journal.pone.0279720}. 
Yet prior work finds that researchers often restrict their use to short-form studies like surveys on such platforms, as more involved methods (e.g., interactive tasks and interviews) are harder to scale and ensure in quality under remote collaborative settings~\cite{spool2001testing,hitlin2016research}. 
Moreover, reliability concerns have grown as participants increasingly misreport backgrounds and use generative AI, estimated at 33–46\% in one study, to complete tasks or game the system for credit~\cite{veselovsky2023artificial, zhang2025generative}.
Instead of being passively undermined by Large Language Models, we can proactively leveraging them.
Initial work has begun exploring simulated user responses for specific tasks: for instance, Perttu et al generated insights by prompting GPT-3 to role-play as a gamer reflecting on gameplay experience~\cite{hamalainen2023evaluating}, and Proxona create simulated personas for audience to help content creators evaluate their work~\cite{10.1145/3706598.3714034}. 
We regard these efforts as locally designed simulated users, an analogue to local recruitment in traditional studies, where simulations are crafted case by case.

Building on this paradigm, we propose shifting from locally crafted simulations to a simulated crowdsourcing procedure. In this approach, AI agents can be “recruited” through pipelines of onboarding, screening, experiencing, and feedback using large profile assets. Such a shift enables scaling up user studies with broader participant diversity, reduced authoring bias, and more generalizable insights. We further highlight the opportunity to build an AI-agent–based crowdsourcing platform, which could reshape user experience research—not as a replacement for traditional methods, but as a complement that provides \textbf{large-scale, low-cost, moderate-fidelity, and good-enough user experiences} to support rapid, early-stage prototyping.

\subsection{LLM Representing Human and the Growing Profile Assets}
There are two types of motivations for using LLM  to represent humans.
One is to establish evaluations akin to Alan Turing’s Imitation Game~\cite{warwick2016can}, where simulating an individual well enough to fool a human judge is regarded as a milestone of artificial intelligence.
This line of work is of diagnostic nature, focusing on the capability gaps between LLMs and humans rather than the utility of simulation~\cite{aher2023using, aharoni2024attributions, binz2023using, hagendorff2022thinking}.

The other is more inclined toward practical value, using LLMs to proxy human behaviors in high-stakes activities, such as social experiments~\cite{park2023generative, park2022social, xu2024ai}, clinical pilots~\cite{natalie2025not, wang2024twin, liu-etal-2025-eeyore}, educational trials~\cite{jin2025teachtune,lu2024generative, markel2023gpteach}, prototype testing~\cite{lu2025uxagent,10.1145/3706598.3714034} and etc, that would otherwise require human participation with potential consequences, but can now be conducted without incurring real-world costs~\cite{aher2023using}.
Accordingly, "profiling LLMs" to achieve accurate simulations of stakeholders in target scenarios constitutes a significant proportion of efforts in this line of research.
For example, Park et al manually author occupation and social relations for 25 generative agents to create a virtual town for social prototyping~\cite{park2023generative}.
Proxona developed an automation pipeline to distill audience traits from comments into personas, helping content creators gain early insights through simulation~\cite{10.1145/3706598.3714034}. 

Ongoing efforts have led to a growing body of profile data assets, ranging from demographic information in population datasets~\cite{Argyle_Busby_Fulda_Gubler_Rytting_Wingate_2023, park2024generativeagentsimulations1000} to domain-derived attributes for profiling specific groups~\cite{natalie2025not, liu-etal-2025-eeyore}, and, more promisingly, large-scale profile materials synthesized by LLMs~\cite{ge2024scaling, park2024generativeagentsimulations1000}.
A notable project profiles generative agents from 1,052 real individuals' audio interviews~\cite{park2024generativeagentsimulations1000}.
An ambitious project releases Persona Hub~\cite{ge2024scaling}, a repository containing 1 billion diverse personas—roughly 1/7 people on Earth~\cite{chen2024persona}.
This progress motivates us to move beyond hypothesis-driven, locally designed simulated users and instead leverage large profile assets to scale up user studies—``recruiting'' and ``screening'' rather than ``inventing''—to overcome the limits of local design (i.e., strong assumptions about what a user should look like) and to gain the benefits of broader perspectives, diversity, and unexpected user types.

\subsection{Critiques on LLM Simulation Validity}
Despite the rapid growth of simulation research, many studies raise critiques of its validity and concerns about risks from inaccurate simulations~\cite{kapania2025simulacrum, schroder2025large}. The main critiques include:
(1) Bias. LLMs inherit biases from their training data, often overrepresenting WEIRD populations while underrepresenting minority groups~\cite{mihalcea2025ai}.
For example,  Rosiana et al find it challenging to use vision–language models to simulate people with low vision~\cite{natalie2025not}.
(2) Limited Diversity. LLMs struggle to capture the variance and diversity of behaviors within groups, even under different persona prompts~\cite{rime2024interviewing, harding2024ai}. For instance, Kwok et al show their failures in simulating cultural diversity~\cite{kwok2024evaluating}.
(3) Construct Validity. Current claims of "successful simulation" face the value–action gap: LLMs may replicate survey answers aligned with human values, but their actions do not necessarily follow these values~\cite{shen2025mind, borah2025mind}.
(4) Robustness. LLMs often fail to consistently react like humans when stimuli are reworded in meaningful but minor ways~\cite{schroder2025large, shu2024you}.
(5) Theoretical Limits. Some argue it is computationally infeasible for models to replicate human psychology across all inputs just based on observations~\cite{van2024reclaiming, harding2024ai}.  Van et al provide a mathematical proof supporting this point of view~\cite{van2024reclaiming}.

We acknowledge these limitations. However, many critiques imply that if simulations are not accurate to capture differences across individuals and groups, they should not be used. But must a successful simulation be perfectly accurate? We respectfully redefine success as a \textit{pragmatically useful simulation} and thus focus on the value of scaling up despite unresolved inaccuracies. Even if fidelity is imperfect at the individual level, aggregated simulated user experiences can still yield holistic insights comparable to those from real users. In our study, we compare insights from simulated agents and human participants, showing that, when we have a way to scale up the user studies, mass simulation produces findings close to those of real users and, as validated by designers, provides actionable guidance for improving prototypes in subsequent design iterations.

\begin{figure}[tb]
\centering
\includegraphics[width=\linewidth]{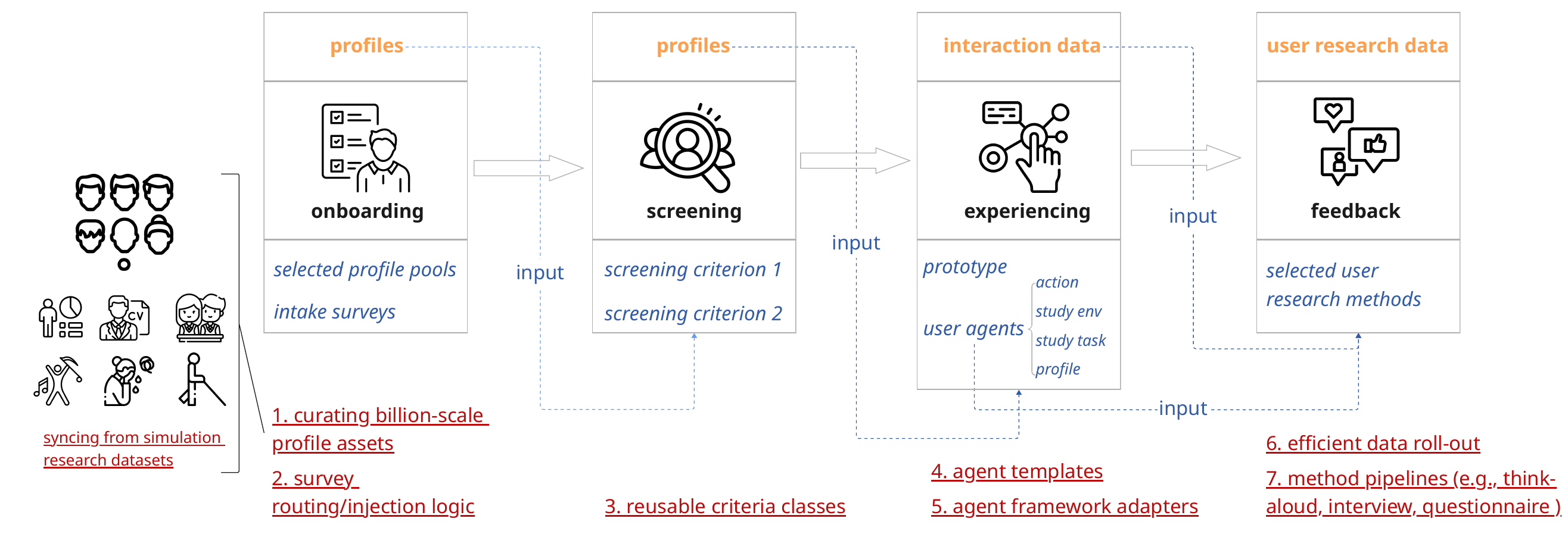}
\caption{Crowdsourcing simulated user agents framework. Each stage produces \textcolor{orange}{outputs}, requires \textcolor{blue}{inputs} with \textcolor{red}{\uline{toolkit supports}}.}
\label{fig:new}
\Description{A pipeline diagram of the Crowdsourcing Simulated User Agents framework, organized into four stages: onboarding, screening, experiencing, and feedback. 
Onboarding takes inputs of selected profile pools and intake surveys, producing enriched profiles with toolkit support such as curating billion-scale profile assets and survey routing logic. 
Screening takes profiles as input, applies criteria (e.g., balance, attribute priority), and outputs user agents, supported by reusable criteria classes. 
Experiencing uses user agents, study environments, study tasks, and action spaces as inputs, producing interaction data, with toolkit supports such as agent templates, framework adapters, and efficient data rollout. 
Feedback takes prototypes, user agents with interaction history, and selected user research methods (e.g., think-aloud, interview, questionnaire) as inputs, producing user research data, supported by method pipelines. 
Toolkit supports are highlighted in red underlined text, inputs in blue, and outputs in orange.}
\end{figure}

\section{Agent Crowdsourcing Framework}
Inspired by human crowdsourcing processes for user experience research, we introduce the idea of \textit{crowdsourcing simulated user agents}, which mirrors four corresponding stages found in human crowdsourcing: onboarding, screening, engaging, and feedback. As illustrated in  Figure~\ref{fig:new}, we design and implement these stages to adapt familiar LLM simulation practices into a crowdsourcing setting. 
The subsections that follow detail each stage: onboarding through intake surveys and profile enrichment (Section~\ref{subsec:surveying}), screening to balance participant pools (Section~\ref{subsec:screening}), experiencing tasks in interactive environments (Section~\ref{subsec:experience}), and feedback through reflective methods such as think-aloud and interviews (Section~\ref{subsec:feedback}). 
We also release a toolkit that supports the construction of this four-stage pipeline.
So we will describe each stage by outlining: what the adapted implementation for this stage is, the underlying design considerations, how it flows in our example game prototyping study, and how it generalizes for UX researchers in broader contexts.

\subsection{Onboarding: Recruiting at Scale with Intake Surveys} \label{subsec:surveying}

\paragraph{Adapted Implementation.} In human crowdsourcing, onboarding refers to the process where participants are brought in, complete intake surveys, and provide basic profile data before they can be screened or assigned tasks. 
In our framework, this stage is implemented by distributing intake surveys to a large number of role-playing LLMs instantiated with basic profiles drawn from existing pools. 
At this point, the role-playing LLMs act as intermediate proxies: they are not yet fully simulated users, since other modules (final profile, environment, action space) are still to be integrated, but they can provide survey responses based on their initial role-play information. 
The outcome is a set of enriched user profiles that combine intake survey results with the basic profiles from the pools.

\paragraph{Design Considerations.} This stage is designed in this way to augment handcrafted profile design with the benefits of scale, broader perspectives, and unexpected user types.  
In our framework, user profiles are constructed from two complementary elements that balance eligibility with diversity.
First, intake surveys provide the \textit{eligibility-driven} component: they allow researchers to encode their recruitment intentions and specify assumptions about what a target user should look like. 
This is the point where handcrafted simulation can be transferred into the crowdsourcing framework, i.e., hypothesized user patterns (e.g., game-playing styles in a gamer study) can be reformulated as survey questions that determine participant eligibility. 
For better understanding, we demonstrate this process in detail in the later example study.
Second, basic profiles provide the \textit{diversity-driven} component: large-scale profile assets from recent simulation research offer auxiliary backgrounds that are not directly relevant to study scenarios but enrich user diversity (e.g., "an extrovert person enjoying jungle adventure"). 
By combining these two components  (e.g., "an extrovert person enjoying jungle adventure with game-playing style x"), the framework scales up and diversifies simulated participants while still respecting researcher-defined eligibility criteria.
A natural question is why we do not simply permute user patterns and auxiliary backgrounds without running LLM inference. 
We think that imitating a human survey process is important: it lets the LLM infer coherent responses that align with the base profile, whereas random permutation risks producing awkward or contradictory combinations (e.g., “a quiet person enjoying parties”), as pointed out by prior work~\cite{li2025llm}.

\paragraph{Illustration Through Example User Study.} We illustrate this process with a game prototyping study in which a team of player participants were recruited to interact with designed NPC prototypes and provide feedback. Details of the NPC prototype design are provided in the Appendix, as they are not the focus here.
For the basic profile pool, we select PersonaHub~\cite{ge2024scaling}, which contains over one billion heterogeneous personas derived from web texts and covering dimensions such as job, hobby, belief, and goal. 
We consider it a strong source for obtaining general and diverse player backgrounds.
We randomly sample 2,900 personas from PersonaHub and prompted GPT-4o~\citep{openai2024gpt4o} to complete two psychometric assessments: the Bartle Test of Gamer Psychology, which categorizes players as Killers, Explorers, Socializers, or Achievers~\cite{bartle2004designing}, and the Big Five Personality Traits test, which characterizes individuals along five dimensions~\cite{goldberg1992development}. 
This process yields 2,900 distinct players ready for the subsequent screening step, completed within a single day.

\paragraph{Application to Broader Research Contexts.} 
To support UX researchers who wish to apply this workflow, our toolkit provides functionality for designing intake surveys and distributing them across curated profile pools.
In particular, we have begun curating one of the largest and most diverse collections of profile resources from existing simulation research. Our current curation spans over a billion large-scale synthetic personas, millions of census-grounded profiles, hundreds of thousands of professional and resume-style profiles, and smaller but high-value stakeholder datasets (e.g., low-vision participants, students). The covered domains are broad, ranging from general purpose to education, accessibility, health, music, and work. Table~\ref{tab:profiles} lists the sources currently included, and we will continue updating and expanding the pools in our toolkit.
By curating across this spectrum, the toolkit abstracts away the low-level complexities of resource gathering, data formatting, and case-specific sampling, and instead offers a unified interface. In practice, researchers can simply select a pool (or several) and begin their study, just as easily as reaching out to participants on a human crowdsourcing platform. To acknowledge original contributions, we properly cite source datasets and require researchers to do the same when they use a given pool.

\subsection{Screening: Applying Criteria to Form Participant Pools} \label{subsec:screening}

\paragraph{Adapted Implementation.} 
In human crowdsourcing, screening ensures that only participants who meet eligibility criteria advance into the study. In our framework, this role is mirrored by applying algorithmic criteria to the enriched profiles produced during onboarding. The inputs are the large set of enriched profiles, and the outputs are a refined pool of user agents that meet researcher requirements. Screening can involve quota-based selection (e.g., equal numbers of player types), curving of attributes to correct for systematic LLM biases, and mechanisms for early stopping when quotas are met. Together, these processes make screening a dynamic stage that shapes the composition of the final participant pool while reducing redundant inference costs.

\paragraph{Design Considerations.} 
The design of this stage reflects the fact that researchers may have different recruitment preferences depending on study goals. Some may prefer balance across traits, while others may prioritize specific attributes. To support both needs, the toolkit allows researchers to specify quotas through a simple dictionary, and participants are heuristically selected once they fall within those quotas. Because it is difficult to predict in advance how many profiles need to be surveyed, we also provide an early-stop mechanism that allows screening to run in parallel with surveying: once quotas are satisfied and the target distribution is achieved, recruitment can be closed immediately, avoiding waste in LLM inference cost. 
Finally, we come up with the idea of using curving tools to address systematic shifts in LLM-generated attributes. 
We have more explanation and demonstration about this point in example user study.

\begin{figure}[ht]
    \centering
    \begin{adjustbox}{valign=t}
    \begin{minipage}{0.48\textwidth}
        \centering
        \text{Before Screening}\\[0.3ex]
        \begin{subfigure}{0.44\textwidth}
            \centering
            \includegraphics[width=\textwidth]{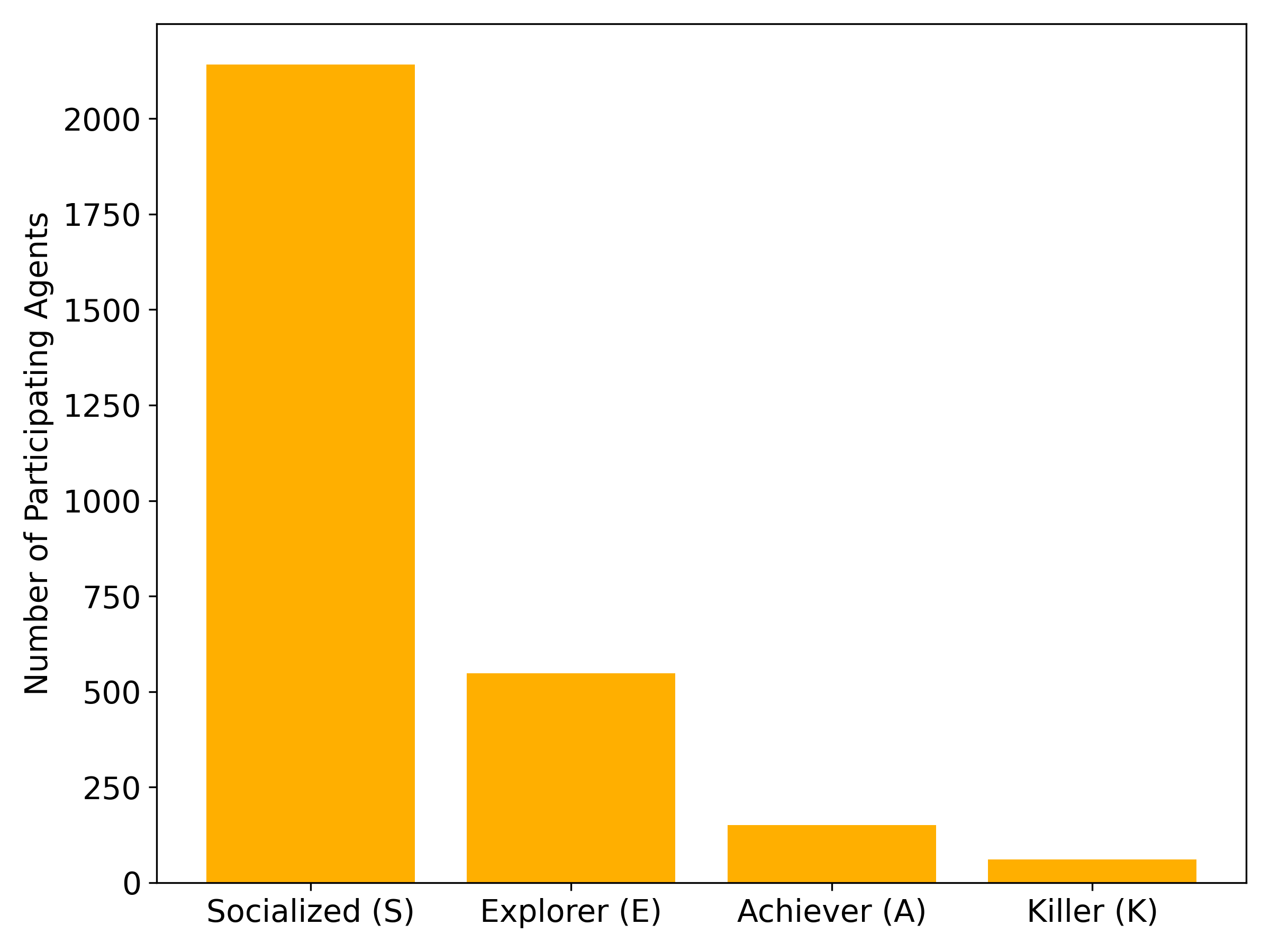}
            \caption{Distribution of the Bartle's Player Types}
            
        \end{subfigure}
        \hfill
        \begin{subfigure}{0.46\textwidth}
            \centering
            \includegraphics[width=\linewidth]{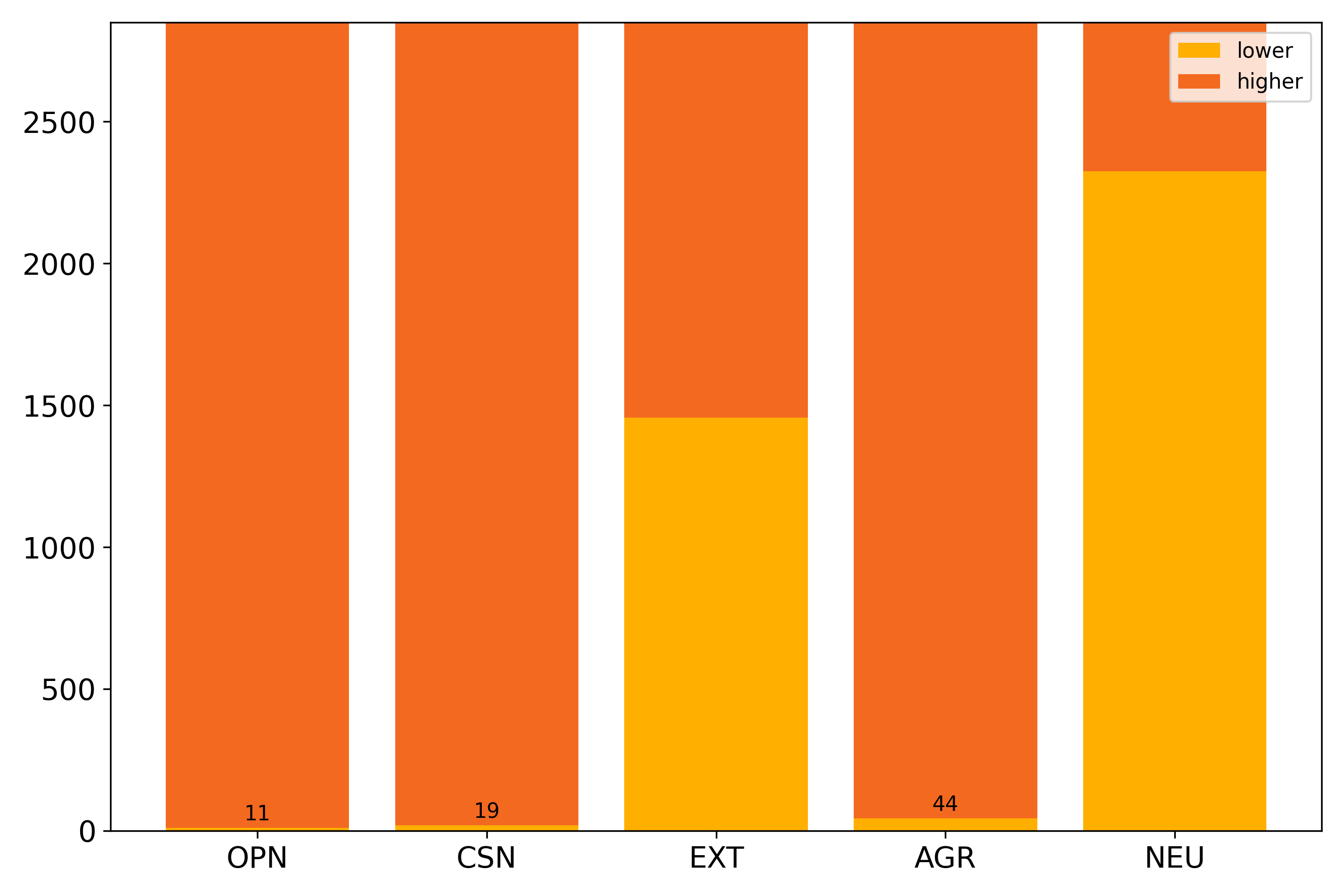}
            \caption{Distribution of Big Five Traits}
        \end{subfigure}
    \end{minipage}
    \hfill
    \begin{minipage}{0.48\textwidth}
        \centering
        \text{After Normalization and Screening}\\[0.3ex]
        \begin{subfigure}{0.44\textwidth}
            \centering
            \includegraphics[width=\textwidth]{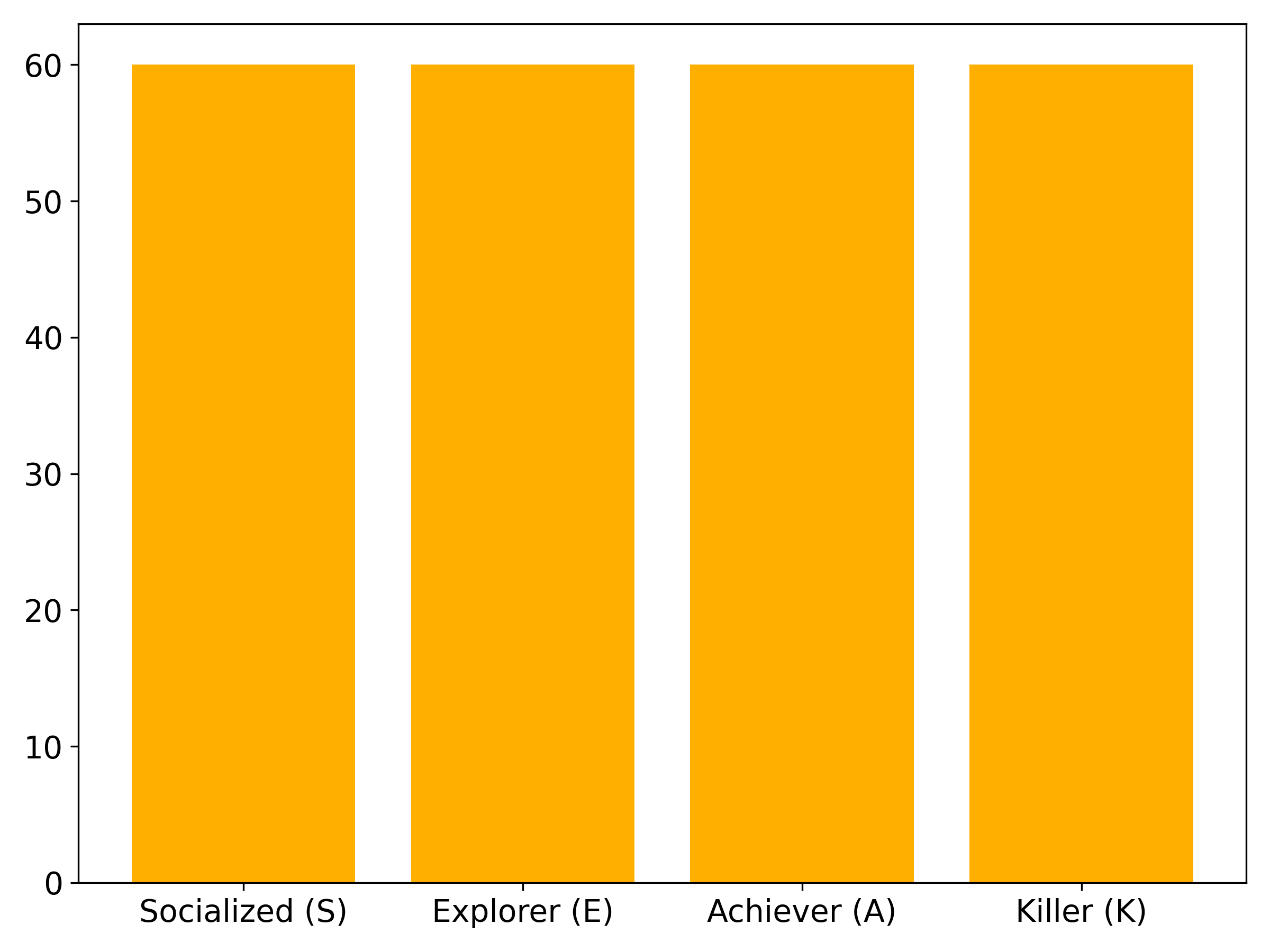}
            \caption{Distribution of the Bartle's Player Types}
        \end{subfigure}
        \hfill
        \begin{subfigure}{0.46\textwidth}
            \centering
            \includegraphics[width=\linewidth]{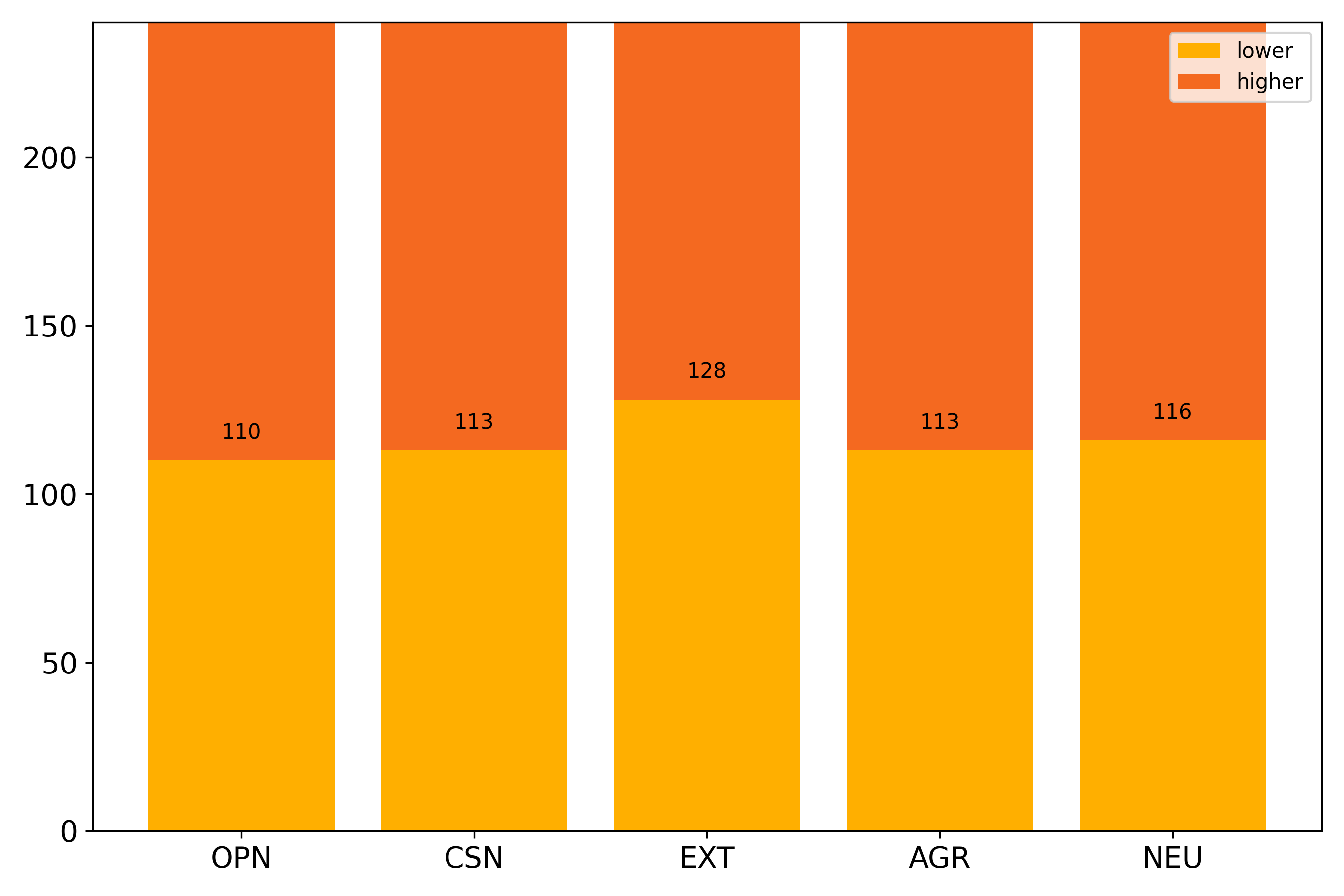}
            \caption{Distribution of Big Five Traits}
        \end{subfigure}
    \end{minipage}
    \end{adjustbox}
    \caption{Comparison of figures before and after calibration and screening, shown in a single row.}
    \label{fig:comparison_row}
    \Description{A side-by-side comparison of participant distributions before and after normalization and screening. 
The left panel, labeled “Before Screening,” shows two subplots: 
(1) the distribution of Bartle’s player types, where Socializers dominate while Killers are scarce; 
(2) the distribution of Big Five traits, showing skewed results with many participants rated as highly open and few with low openness, conscientiousness, or high neuroticism. 
The right panel, labeled “After Normalization and Screening,” shows two corresponding subplots: 
(1) the Bartle’s player types distribution is balanced across Killers, Socializers, Explorers, and Achievers; 
(2) the Big Five traits distribution is normalized, with a more even spread across low and high trait levels. 
This demonstrates how screening corrected initial imbalances to form a more representative participant pool.}

\end{figure}

\paragraph{Illustration Through Example User Study.}
During surveying, we continuously monitor the distribution of survey results so that we could stop recruitment as soon as a balanced team was achievable. 
After surveying 2,900 profiles, we assemble a satisfying team of 240 player profiles. 
Early inspection show that the raw distribution of player types and Big Five traits was highly imbalanced.
For example, there are significantly more Socializers than Killers, and highly open individuals are far more than those with lower openness (see Figure~\ref{fig:comparison_row}, left). So, we need do ``screening''.
This imbalance not only reflects natural imbalances in human populations (e.g., socializers are more than killers), but also is due to a behavioral tendency of GPT: it tends to avoid assigning low openness, low conscientiousness, or high neuroticism, likely because LLMs are trained to be supportive and kind, and avoid attributing traits that may seem negative.
So, following prior work addressing biases in LLMs’ psychological assessments, we apply "curving" - a normalization process - to the Big Five scores. 
For instance, openness greater than the group average score 4.69 (out of 5) is considered high openness, while scores below 4.69 is interpreted as lower openness.
After Big Five Trait normalization and careful selection, we form a team of 240 player agents with balanced distributions, including all kinds of Bartle player types and diverse levels of Big Five traits (see Figure~\ref{fig:comparison_row}, right).

\paragraph{Application to Broader Research Contexts.}
In broader applications, the screening stage generalizes to any scenario where researchers need to form targeted participant pools. 
The toolkit supports this process by offering reusable criteria classes for both balance-first and priority-first selection, enabling flexible alignment with study goals. 
Because surveying and screening can run simultaneously, researchers can monitor distributions in real time and stop the recruitment process as soon as quotas are met. 
This greatly reduces inference costs while ensuring that the final team reflects researcher intentions. The toolkit also makes curving tools available for correcting systematic LLM biases, helping researchers avoid overrepresentation of certain traits while preserving diversity. 
Together, these features make screening a powerful stage that transforms raw profile data into a curated pool of simulated participants.

\vspace{-0.5em}

\subsection{Experiencing: Simulated Interactions with Study Tasks}\label{subsec:experience}

\paragraph{Adapted Implementation.} 
In human crowdsourcing, participants engage with study environments and tasks, producing interaction logs and behavioral data.
In our framework, this stage is mirrored by simulated participants interacting with task environments through structured prompts. 
In our framework, this stage is mirrored by simulated participants interacting with prototypes or tasks through structured prompts. 
The inputs are the established simulated user agents, enriched with participant profiles, study environments, study tasks, and an action space, as well as the prototype they engage with. 
The outputs are detailed records of interactions—including dialogue, actions, and task outcomes—that serve both as a memory for coherent feedback and as behavioral traces for subsequent analysis.

\paragraph{Design Considerations.} 
Unlike the first two stages, which focus only on constructing and refining profiles, this stage requires the creation of fully instantiated agents capable of acting within an environment. 
The goal is to move beyond static reflections or post-hoc opinions, which prior simulation work often relied on~\cite{hamalainen2023evaluating}, and instead enable real-time, dynamic engagement with study prototypes. 
To achieve this, the agent architecture integrates multiple modules such as environment, profile, goal, memory, and action space, so that simulated users can behave coherently and contextually. 
Embedding the study-specific details into these modules ensures that agents interact in ways that are faithful to the scenario and useful for design evaluation.

\paragraph{Illustration Through Example User Study.}
In our game prototyping study, we define key modules directly within agent prompts. 
As shown in the system prompt of player agents in Figure~\ref{fig:npc_interaction}, each agent was instructed with background details (e.g., PersonaHub profile, Bartle type, Big Five traits), in-game character roles, task goals, and a defined action space, which are author-prepared information for this user study. 
Agents then engage in immersive interactions with NPCs one by one through a text-based interface. During the interaction, both the agent and the NPC selected actions each round (see Table~\ref{tab:action_definitions}). 
Interactions conclude either when the player agent selected the \texttt{[D-END]} action, indicating that the goal has been reached, or after 30 turns had elapsed. 
This yields transcripts that captured not only outcomes but also decision-making processes and context-sensitive behaviors.

\paragraph{Application to Broader Research Contexts.}
To make this stage broadly reusable, our toolkit abstracts the agent-building process into templates that researchers can adapt to their own domains. These templates provide guidance on how to embed study environment information into prompts, define action spaces tailored to task scenarios, and control interaction length and outcomes. 
The toolkit also supports integration with existing agent frameworks such as AutoGen and compatibility with different LLM backends, including Gemini and OpenAI models. 
This flexibility allows researchers to plug in their own study environments and models while benefiting from a standardized pipeline for generating scalable, analyzable interaction data.

\subsection{Feedback: Eliciting Participant Reflections and Evaluations}\label{subsec:feedback}

\paragraph{Adapted Implementation.} 
In human crowdsourcing, participants are often asked to provide reflective feedback through user research methods such as surveys, interviews, or think-aloud protocols. In our framework, the inputs are simulated user agents together with their recorded interaction histories and the chosen research instruments. The outputs are structured reflections and evaluations generated by the agents. Feedback can be elicited both during tasks and after completion, for example, through turn-level think-aloud logs or post-task questionnaires and interviews. This stage consolidates experiential data into higher-level insights that can be analyzed alongside human feedback.

\paragraph{Design Considerations.} 
The central aim of this stage is to support the full repertoire of user research methods, adapting them from human participants to simulated agents. 
Prior work has largely focused on simpler uses, such as prompting simulated agents to answer survey questions or usability checklists. 
While such approaches are valuable for validating simulation fidelity in controlled scenarios, they miss the richer, subjective expressions that make human feedback so valuable. 
Our framework seeks to extend beyond survey responses by enabling think-aloud protocols, interviews, and, eventually, even more qualitative methods such as ethnographic elicitation. 
By supporting this variety, the toolkit allows researchers not only to collect quantitative data but also to probe simulated agents for deeper, more nuanced perspectives that parallel traditional human-centered methods.

\begin{figure}[t]
    \centering
    \includegraphics[width=\linewidth]{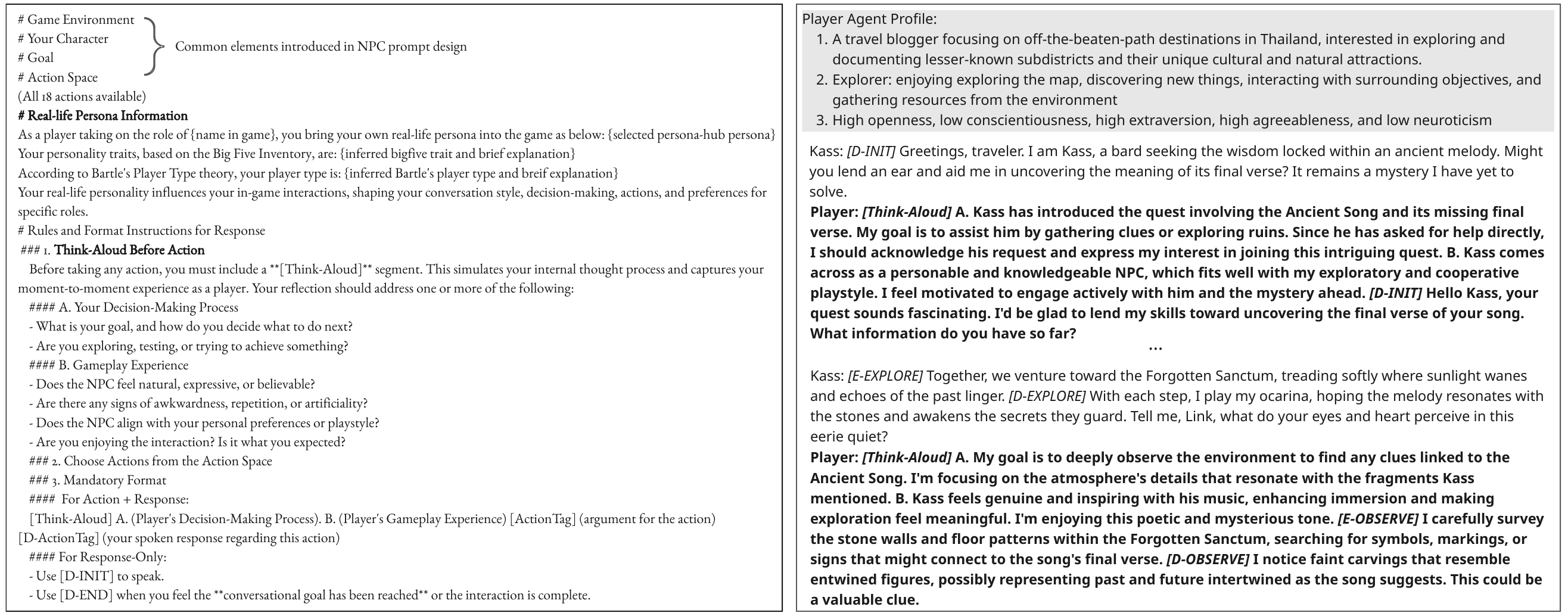}
    \caption{Example interaction with NPC Kass. Left: player agent prompt; Right: dialogue with think-aloud and actions.}
    \Description{The figure shows an example of a simulated player agent interacting with the NPC Kass. 
On the left side is the system prompt given to the player agent, which includes background details and instructions. 
On the right side is a dialogue exchange: each turn contains a think-aloud segment where the agent explains its reasoning, 
followed by the selected action (e.g., initiating dialogue or accepting a quest), and then the corresponding spoken response. 
This illustrates how think-aloud protocols and action choices are embedded in simulated interactions.}

    \label{fig:npc_interaction}
    \vspace{-1.5em}
\end{figure}

\paragraph{Illustration Through Example User Study.} 
In our game prototyping study, simulated user agents were asked to perform think-aloud during interactions and to participate in an interview after completing all tasks. The think-aloud segments and interaction logs serve as a form of memory, grounding the subsequent interview responses in the agent’s prior actions and reflections. We assume that incorporating think-aloud, similar to a self-reflection prompting strategy, helps simulated participants generate more coherent and contextually grounded feedback during interviews.
\subparagraph{Think-aloud Protocol.}
To capture the agent's internal reasoning and experience in the interaction as a player, we implement a structured \textit{think-aloud protocol}. Player agents are instructed to generate a \texttt{[Think-Aloud]} segment before taking any action at each turn. 
This segment serves to simulate a moment-by-moment reflection on (1) their decision-making process and (2) game-play experience, mirroring traditional think-aloud methods used in human-centered research.
The think-aloud protocol is embedded in the player agent’s system prompt. An example of an agent response that includes the think-aloud, chosen action, and response is shown in Figure~\ref{fig:npc_interaction}.

\subparagraph{Interview.}
We conduct semi-structured interviews after players complete all eight NPC interactions. This interview is designed to gather holistic feedback on the quality and experience of interacting with LLM-driven NPCs. The questions targeted nine key aspects, ranging from language authenticity to personal fit, with particular attention to areas not easily captured through automated metrics.
The full set of interview prompts is listed in Table~\ref{tab:interview}.
These interviews allow us to better understand how different players perceived the system's responsiveness, immersion, and alignment with their play styles. 

\paragraph{Application to Broader Research Contexts.}
To broaden applicability, our toolkit provides reusable modules for implementing multiple feedback methods, including think-aloud protocols, structured questionnaires, and semi-structured interviews. Researchers can select a method that best aligns with their study goals and scale requirements: for example, lightweight questionnaires for thousands of participants, or richer interview-style feedback for smaller but more in-depth studies. 
Because feedback generation is decoupled from real-time interaction, it can be executed in batch mode, enabling thousands of agents to be “interviewed” in parallel without sequential overhead. 
The toolkit further integrates multi-worker processing and support for common API providers (e.g., OpenAI, Anthropic, Gemini), reducing both runtime and costs. This flexibility makes it possible to scale qualitative-style methods to sizes that are infeasible with human participants, while still retaining the richness of structured reflection.

\section{Evaluation}
With the finally obtained agent simulated user study, we conduct three alternative user studies.
First, we recruit 10 students from the authors' university who have experience in related games we test. 
Second, we recruit 20 participants from the Prolific crowd-sourcing platform who identifies video gaming as a hobby and has experience with the games we test. 
Demographic and compensation information of both human user study are illustrated in Appendix~\ref{appx:human}.
Lastly, we create a basic agent using GPT-4.1-mini, with just specifying a basic player role. This model serves as a generic participant in the evaluation.


\subsection{Validating the Effect of Scaling}
\begin{wrapfigure}{r}{0.4\textwidth}
    \centering
    \scalebox{1}{
    \includegraphics[width=0.8\linewidth]{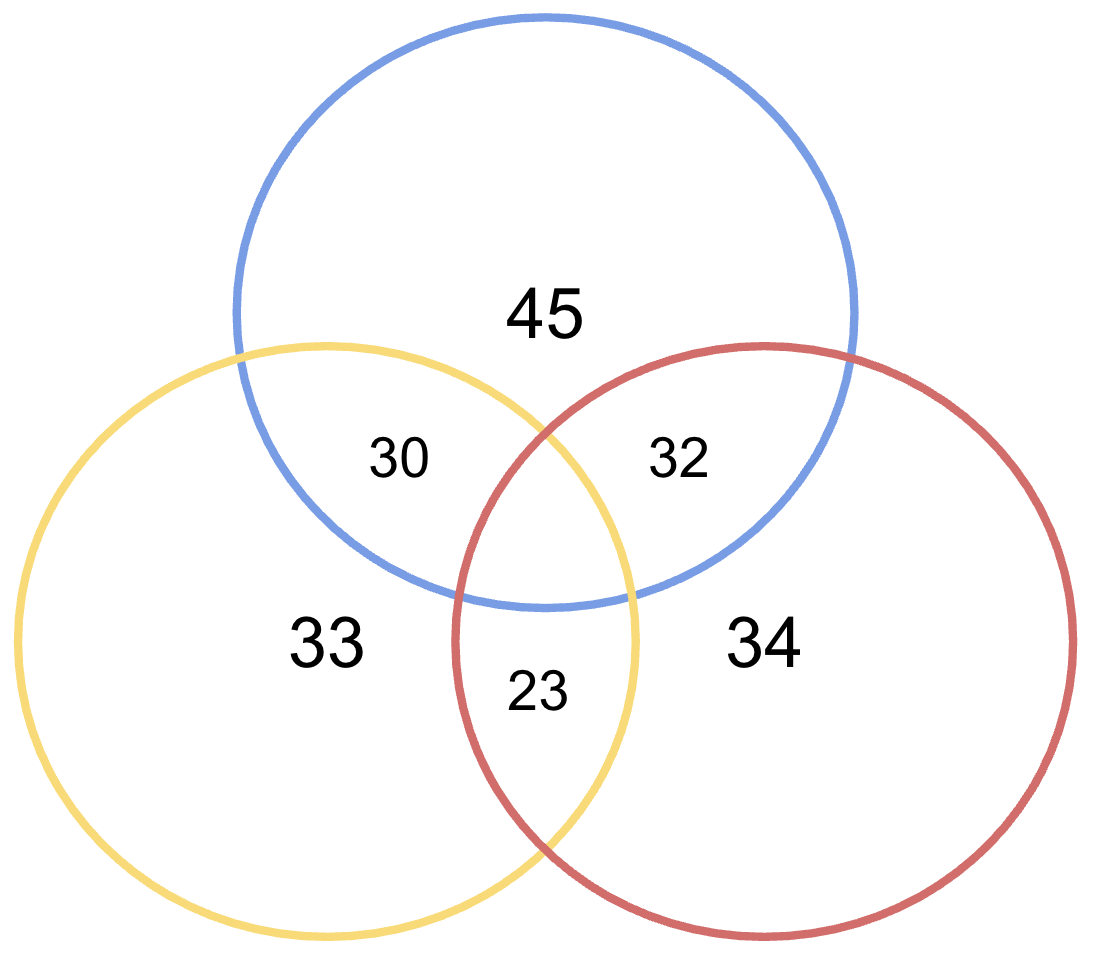}}
    \caption{Overlap of qualitative codes across three user studies: 10 local human participants (yellow), 20 crowdsourced human participants (red), and 240 simulated agent participants (blue).}
    \Description{A Venn diagram showing the overlap of qualitative codes identified across three user studies. 
Yellow represents 10 local human participants, red represents 20 crowdsourced human participants, and blue represents 240 simulated agent participants. 
The diagram highlights areas of overlap, indicating which insights were unique to each group and which were shared across groups.}

    \label{fig:overlap}
\end{wrapfigure}

\paragraph{Coding Procedure.} To examine the relationship between scaling and simulation usefulness, we first interpret usefulness as the extent to which agent transcripts reproduce findings observed in human interviews. The more human-derived findings that also appear in agent transcripts, the more valuable the agents are for generating actionable design insights.
We apply inductive coding to the transcripts of the crowsourced agent study and the two human studies.
Two authors independently code each set of transcripts and resolved disagreements through discussion.
Because coding is conducted separately for each study, semantically similar codes sometimes appeared under different labels. 
To unify them, we first consolidate overlapping codes across the two human studies by reviewing their transcripts together. 
Codes judged to reflect the same phenomenon are merged, while keeping unmatched codes distinct. 
We then compare the resulting human union set against the agent codes, repeating the consolidation process. 
This yields a final unified codebook of 48 codes spanning all three studies.
This yields a final union of 48 codes across all three studies. 
Figure~\ref{fig:overlap} visualizes the overlap among codes across the three studies, and Table~\ref{tab:freq_least} highlights the five most frequent and five least frequent codes in agent transcripts with representative examples.

\paragraph{Subsampling agent teams.} To assess whether high coverage arises primarily from the scale of the agent team, we estimate how coverage changes as team size decreases. 
Rather than re-running new studies with fewer agents, we resample from the existing 240-agent transcripts. 
Because codes are fixed once transcripts are coded, we can simply recalculate the number of distinct codes recovered from smaller subsets of agents. 
Specifically, we sample agent transcripts at team sizes of 1, 2, 4, \ldots, 128 (up to the maximum of 240), repeating each sampling 10 times to smooth variation. 
For each team size, we compute the average fraction of human codes recovered, i.e., coverage of human findings. 
This subsampling procedure allows us to model how coverage scales with the number of agents and to compare it against the human baselines.

\begin{table}[tb]
\centering
\small
\caption{Top 5 most frequent and least frequent qualitative codes in the agent interview transcripts with representative example.}
\label{tab:freq_least}
\begin{tabular}{p{4cm} p{0.6cm} p{9.4cm}}
\toprule
\multicolumn{3}{c}{\textbf{Top 5 Most Frequent Codes}} \\
\textbf{Code Name} & \textbf{Freq.} & \textbf{Example} \\
\midrule
Redundant response / repetition & 518 & Some lines felt slightly repetitive or mechanical, which pulled me out of immersion. \\
Immersive experience & 475 & NPCs remembered earlier points and advanced the narrative organically. \\
Clear goals & 389 & Each interaction had clear goals, such as accepting a quest or uncovering lore. \\
Smooth system flow & 376 & NPCs reliably responded and maintained context. \\
Appropriate response & 307 & NPCs stayed relevant to context and built logically on prior discussion points. \\
\midrule
\multicolumn{3}{c}{\textbf{Top 5 Least Frequent Codes}} \\
\textbf{Code Name} & \textbf{Freq.} & \textbf{Example} \\
\midrule
Confused by the low-fidelity setting & 2 & The range of options felt constrained, likely due to low-fidelity NPC design. \\
Lengthy response & 2 & I might prefer more concise dialogue and less verbose exchanges. \\
Unnecessary / over freedom & 5 & Too much openness sometimes felt overwhelming or unfocused. \\
Information overload & 7 & The wide range of options could sometimes feel overwhelming. \\
Off topic & 8 & There were minor moments where NPC replies seemed slightly off-topic. \\
\bottomrule
\end{tabular}
\end{table}

\begin{figure}[tb]
  \centering
  \scalebox{0.7}{
  \includegraphics[width=\linewidth]{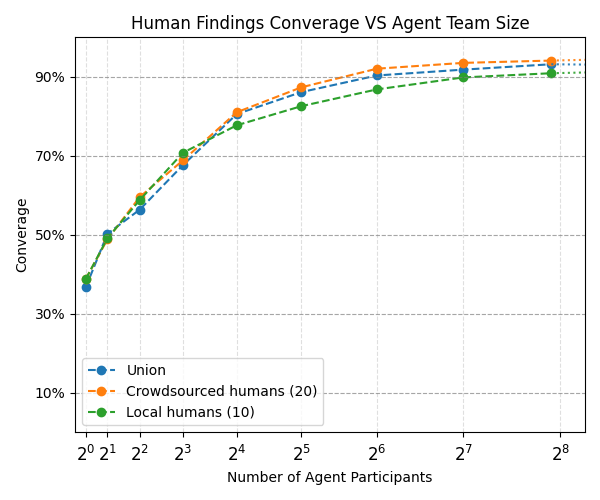}}
  \caption{Coverage of human study findings increases smoothly as we increase the number of simulated user agents. 
  Coverage is defined as the fraction of distinct \textit{codes} identified in human transcripts that are also recovered in agent transcripts, i.e., $\text{coverage} = \frac{|\{ C_\text{human} \cap C_\text{agent} \}|}{|\{ C_\text{human} \}|}$, where $C_{human}$ is the set of distinct \textit{codes} in human transcripts.}
  \label{fig:scaling}
  \Description{A line chart showing the relationship between the number of simulated user agents and the coverage of human study findings. 
The x-axis represents team size of simulated agents, sampled from 1 up to 240. 
The y-axis represents coverage, defined as the fraction of distinct codes identified in human transcripts that are also recovered in agent transcripts. 
The curve rises smoothly as team size increases, then plateaus at around 90\% coverage, indicating diminishing returns after large numbers of agents.}

\end{figure}

\paragraph{Results}
Figure~\ref{fig:scaling} shows a clear scaling effect: as the number of simulated agents increases, the coverage of human-derived findings rises smoothly and plateaus at around 90\%. 
The plateau indicates a persistent gap from 100\%, meaning that even if useful, simulated studies cannot fully substitute for traditional human studies.
We also observe a small but consistent difference between the two human baselines. 
Coverage of crowdsourced human findings reaches 90\% with a team of about 64 agents, whereas achieving the same level for local human findings requires about 128 agents. 
This gap is expected, since the local study likely produced richer and deeper insights due to closer researcher supervision and higher process quality, making them harder for agents to fully replicate.  
By interpreting these scaling curves, we make these interesting equivalencies: 12.8 simulated
agents are as useful as one locally recruited human, and 3.2 agents are as useful as one crowdsourced human.
It is important to emphasize, however, that this effect should not be mistaken for a universal ``scaling law'' of simulation. 
The ratios observed here depend on the complexity and context of the user study. 
In more demanding settings, more agents may be needed to approximate a single human, whereas in more intuitive tasks the ratio could be smaller.  
Overall, the findings demonstrate that while individual agents are imperfect, aggregating them at scale produces insights that are representative and comparable to those from traditional user studies.

\subsection{Expert Evaluation}

\subsubsection{Procedure}

\begin{table}[ht]
\centering
\caption{
Comparison of time and cost across user study configurations. Time reported in minutes. Inter./P = NPC interactions per player; Cost/P = cost per player; Total/P = total time per player.
}
\label{tab:combined_time_cost_clean}
\scalebox{0.65}{
\begin{tabular}{lcc|cccc|ccc}
\toprule
\textbf{Team} & \textbf{Size} & \textbf{Inter./P} 
& \multicolumn{4}{c|}{\textbf{Time Consumption}} 
& \multicolumn{3}{c}{\textbf{Cost}} \\
\cmidrule(lr){4-7} \cmidrule(lr){8-10}
& & & \textbf{Recruit} & \textbf{Interact} & \textbf{Post-interact} & \textbf{Time/P} 
& \textbf{Cost/P} & \textbf{Cost/Insight} & \textbf{Note} \\
\midrule
Agentic         & 240 & 8 & 240   & 1380   & 120   & 6.9   & \$0.28  & \$6.03   & API-call \\
Crowdsourced    & 20  & 1 & 402   & 600    & 315   & 65    & \$20.50 & \$31.53  &compensation \\
Local     & 10  & 1 & 498   & 300    & 150   & 95    & \$40.00 & \$33.33  & compensation \\
LLM-as-generic-user       & 1   & 8 & 0     & 5.4    & 0.5   & 5.9   & \$0.14     & \$0.028      & Minimal API-call \\
\bottomrule
\end{tabular}}
\end{table}

To evaluate practical trade-offs across user study configurations, we locally recruit three experienced game developers to serve as expert reviewers. They assess the four study setups on four dimensions: time efficiency, cost efficiency, fidelity, and insight helpfulness.
Experts are asked to immerse themselves in a game NPC prototyping scenario, seeking feedback from users for iterative development.
Within this framing, we offer four user studies' outcomes, and the experts, acting as adopters, evaluate each study configuration with consideration of budget and time constraints while aiming to get meaningful and actionable player insights.
For each configuration, we provided a packet containing: (i) interaction logs, (ii) interview transcripts, (iii) detailed time and cost breakdowns (briefly summarized in Table~\ref{tab:combined_time_cost_clean}), and (iv) the set of coded insights and design implications derived from the interviews (see Table~\ref{tab:llm_npc_insights} for an example). Insights and implications were produced by two authors via inductive coding of the four studies’ interview transcripts.
We recruit three experts from the University through distribution by the computer science department coordinator, with compensation of 60\$ per hour. 
Participants are required to have significant experience in game design and development, such as completing a game design course. 
Their relevant experience is confirmed through brief verbal interviews.
After reviewing the packet for each configuration, experts rated it on four dimensions using the following rubric:

\begin{wrapfigure}{r}{0.45\textwidth}
\centering
    \centering
    \small
    \captionof{table}{Evaluation scores from three game experts across four user studies.}
    \scalebox{0.7}{
    \begin{tabular}{lcccc}
    \toprule
    \textbf{Dimension} & \textbf{Agentic} & \textbf{Local} & \textbf{Crowdsourcing} & \makecell{\textbf{LLM-as-}\\ \textbf{generic-user}} \\
    \midrule
Expert 1 & 5.00 & 1.00 & 2.00 & 5.00 \\
Expert 2 & 5.00 & 1.00 & 2.00 & 5.00 \\
Expert 3 & 5.00 & 1.00 & 2.00 & 5.00 \\
\midrule
\multicolumn{5}{l}{\textit{Cost Efficiency (rated by experts)}} \\
Expert 1 & 5.00 & 1.00 & 3.00 & 5.00 \\
Expert 2 & 4.00 & 2.00 & 3.00 & 5.00 \\
Expert 3 & 4.00 & 2.00 & 2.00 & 5.00 \\
\midrule
\multicolumn{5}{l}{\textit{Fidelity (computed)}} \\
Expert 1 & 3.42 & 4.06 & 4.06 & 2.24 \\
Expert 2 & 2.94 & 4.20 & 3.70 & 1.62 \\
Expert 3 & 2.58 & 4.09 & 3.84 & 1.82 \\
\midrule
\multicolumn{5}{l}{\textit{Insight Helpfulness (NDCG)}} \\
Expert 1 & 3.53 & 3.71 & 2.82 & 0.43 \\
Expert 2 & 3.88 & 3.22 & 3.55 & 2.36 \\
Expert 3 & 4.00 & 5.00 & 4.68 & 3.10 \\
    \bottomrule
    \end{tabular}}
    \label{tab:expert_evaluation}
\end{wrapfigure}

\paragraph{(1) Time:} Experts receive details on time spent in recruitment, interaction, and post-interaction taks (summarized in Table~\ref{tab:combined_time_cost_clean}). They rate time efficiency on a 1–5 scale, where 5 = $\ge$5× faster than the baseline (Local Participants), 4 = 4–5×, 3 = moderately efficient, 2 = marginally faster, and 1 = comparable/slower, treating local participant study as baseline.
\paragraph{(2) Cost:} Experts review the budget breakdown for each configuration (also summarized in Table~\ref{tab:combined_time_cost_clean}) and rate cost efficiency on a 1–5 scale: 5 = highly efficient (optimal ROI), 4 = relatively efficient, 3 = moderately efficient, 2 = over budget (needs improvement), and 1 = extremely inefficient ($>$10× over budget), treating local participant study as baseline.
\paragraph{(3) Fidelity:} Fidelity is assessed via player behavior and insight fidelity. For behavior fidelity, experts first experience the NPCs and then review Local Player transcripts to identify five representative behaviors (e.g., “building rapport early in interaction”). They also propose five additional behaviors expected from typical players. For each study, transcripts are checked for the presence of these 10 behaviors, scoring 1 for each observed. Local participants are expected to achieve full recall for their half. Behavior fidelity is computed per expert as:
{\small
\[
\text{BehaviorFidelity}_i = \frac{1}{3} \sum_{j=1}^{3} \frac{b_{i,j}}{10},
\]
}

where
$b_{i,j}$ denotes the number of matched behaviors in expert $j$'s defined behavior group observed in study $i$.

For insight fidelity, we use the insights from the Local and Crowdsourced groups as the reference sets. Experts identifies overlapping insights between each study and these two references, and we compute the Jaccard index~\cite{jaccard1912distribution}:
{\small
\[
\text{InsightFidelity}_i = \frac{1}{2}(\frac{|\text{Study}_i \cap \text{Local Human}|}{|\text{Study}_i \cup \text{Local Human}|}  +  \frac{|\text{Study}_i \cap \text{Crowdsourced Human}|}{|\text{Study}_i \cup \text{Crowdsourced Human}|})
\]
}

The final fidelity score is the average of behavior fidelity and insight fidelity and normalized into range from 1 to 5:
{\small
\[
\text{Fidelity}_i = \frac{\text{BehaviorFidelity}_i 
 + \text{InsightFidelity}_i}{2} * 5
\]
}

\paragraph{(4) Insight Helpfulness:} Experts review all insights from the four studies and select the 10 most useful ones for NPC design with importance in order. For each ranked insight, they indicate its source(s) of study. Insight helpfulness is scored via NDCG~\citep{jarvelin2017ir}, a metric accounting for ranked importance:
{
$
\text{InsightHelpfulness}_i = \sum_{k=1}^{10} \frac{\text{presence}_{k,i}}{\log_2(k + 1)} * 5,
$
}
where $presence_{k,i}$ = 1 if the $k$-th important insight included study $i$ as a source.


\subsubsection{Results}




\begin{wrapfigure}{l}{0.45\textwidth}
    \centering
    \includegraphics[width=0.8\linewidth]{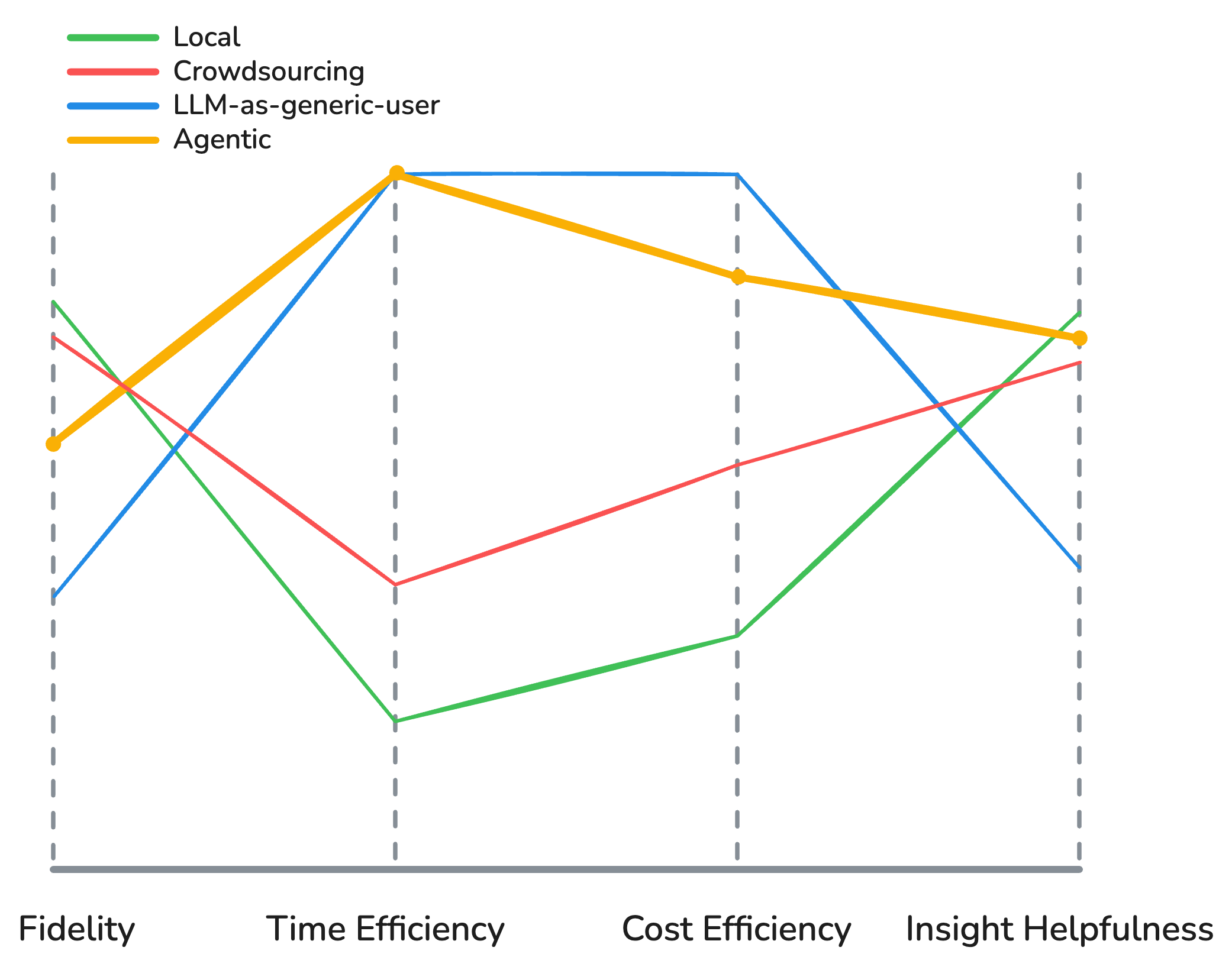}
    \caption{Average expert ratings.}
    \Description{A radar chart showing average expert ratings across four study configurations: Agentic, Local, Crowdsourcing, and LLM-as-generic-user. 
Each axis corresponds to a dimension: time efficiency, cost efficiency, fidelity, and insight helpfulness. 
The Agentic method scores highest on time and cost efficiency, with moderate fidelity and insight helpfulness. 
Local participants score highest on fidelity and insight helpfulness but lowest on time and cost. 
Crowdsourcing scores in between, with moderate ratings across all dimensions. 
The LLM-as-generic-user scores high on time and cost but low on fidelity and insight helpfulness.}
    \label{fig:pareto}
\end{wrapfigure}

Figure~\ref{fig:pareto} visualizes the average
expert rating for four study configurations
across all dimensions. Table~\ref{tab:expert_evaluation} shows how experts rated each user study method. Agreement among the three experts was strong (ICC(2,1) = 0.817, 95\% CI [0.640, 0.920]).
From the results, we see that the agentic player team is highly efficient in both time and cost, while still offering solid performance in fidelity and insight helpfulness. Local participants achieved the highest scores in fidelity and insight quality but incurred much higher time and budget costs. Crowdsourced players performed reasonably well but remained slower and more expensive. The LLM-as-generic-user was fast and cheap but struggled to deliver meaningful insights or maintain fidelity. 
Overall, our evaluation confirms the potential of agentic AI to generate low-cost, moderate-fidelity, yet good-enough user experience in this early-stage prototyping scenario.
We conjecture that traditional methods may still be better suited for capturing rich, high-fidelity insights in later stages of design, and we leave this for future exploration.

\section{Conclusion}
This work has introduced \textit{Crowdsourcing Simulated User Agents}, a scalable method and toolkit that adapts familiar human crowdsourcing logic, onboarding, screening, experiencing, and feedback, to the context of large-scale LLM-based agents.  
Through a game prototyping study, we demonstrated how this method can transform billions of profile assets into hundreds of simulated participants, producing insights that professional designers rated as balancing fidelity, cost, time efficiency, and usefulness. Our findings highlight a clear scaling effect: while individual agents are imperfect, aggregating them yields representative and actionable insights comparable to those from human participants, at a fraction of the time and cost.  
We position this framework not as a replacement for human studies, but as a complementary option that extends the UX researcher’s toolkit.

\begin{acks}
To Robert, for the bagels and explaining CMYK and color spaces.
\end{acks}

\bibliographystyle{ACM-Reference-Format}
\bibliography{sample-base, custom}

\appendix

\begin{table}[h]
\centering
\scalebox{0.9}{
\small
\begin{tabular}{p{3cm} p{4.5cm} l l l l l}
\toprule
\textbf{Name} & \textbf{Intro} & \textbf{Domain} & \textbf{Size} & \textbf{Attribute} & \textbf{Format} & \textbf{Source} \\
\midrule
US Census Sampler~\cite{chang2025llms} & Profiles sampled from the joint distribution of age, gender, and race/ethnicity in U.S. census data. & general & sampled & objective & structured & real-world \\
PersonaHub~\cite{ge2024scaling} & Large-scale heterogeneous personas derived from web text, covering dimensions such as job, hobby, belief, goal, etc. & general & billions & mix & descriptive & synthesized \\
Tianyi-Personas~\cite{li2025llm} & Synthesized personas  with census data as demographic grounding. & general & millions & mix & structured & synthesized \\
Persona-Chat~\cite{zhang-etal-2018-personalizing} & Short four-sentence self-descriptions with personal chats crowdsourced from participants. & general & 1,115 & subjective & descriptive & real-world \\
Low-vision~\cite{natalie2025not} & Vision perception profiles of participants with impaired vision. & accessibility & 40 & objective & structured & real-world \\
Student~\cite{jin2025teachtune} & Expert-defined traits characterizing students in educational contexts. & education & 243 & objective & structured & expert-derived \\
talkpl~\cite{talkpl_userprofile_2025} & Music user profiles including preferences and listening habits. & music & 33k & objective & structured & real-world \\
Eeyore~\cite{liu-etal-2025-eeyore} & Extracted profiles of individuals with depression-related traits from conversations. & mental health & 2,589 & mix & structured & expert-derived \\
Djinni~\cite{drushchak-romanyshyn-2024-introducing} & Job-seeker profiles collected from an online recruitment platform. & profession & 230k & objective & descriptive & real-world \\
Resume~\cite{salvador_resume_extensive_dataset_2025} & Resume-style profiles for job-seeking individuals. & profession & 590k & objective & structured & real-world \\
\bottomrule
\end{tabular}}
\caption{Available profile assets we are currently curating.}
\label{tab:profiles}
\end{table}

\begin{table}[h]
\centering
\scalebox{0.7}{
\begin{tabular}{|l|p{30em}|}
\hline
\textbf{Action Type} & \textbf{Definition} \\ \hline
\textbf{D-INIT} & Speaking: Initiating or continuing a conversation. Format: [D-INIT] (your text) \\ \hline
\textbf{D-END} & Ending a Conversation: Concluding a conversation. Format: [D-END] (your text) \\ \hline
\textbf{Q-ACCEPT} & Accepting a Quest: Agreeing to take on a quest. Format: [Q-ACCEPT] (quest description) [D-ACCEPT] (response) \\ \hline
\textbf{Q-REJECT} & Rejecting a Quest: Declining a quest. Format: [Q-REJECT] (quest description) [D-REJECT] (response) \\ \hline
\textbf{Q-OFFER} & Offering a Quest: Proposing a quest. Format: [Q-OFFER] (quest description) [D-OFFER] (response) \\ \hline
\textbf{Q-COMPLETE} & Completing a Quest: Fulfilling quest requirements. Format: [Q-COMPLETE] (completion confirmation) [D-COMPLETE] (response) \\ \hline
\textbf{E-OBSERVE} & Observing Details: Looking for clues. Format: [E-OBSERVE] (description) [D-OBSERVE] (response) \\ \hline
\textbf{E-INTERACT} & Interacting with an Object: Engaging with an object. Format: [E-INTERACT] (description) [D-INTERACT] (response) \\ \hline
\textbf{E-EXPLORE} & Exploring a Location: Investigating a new area. Format: [E-EXPLORE] (location) [D-EXPLORE] (response) \\ \hline
\textbf{E-GATHER} & Gathering Resources: Collecting items. Format: [E-GATHER] (resources) [D-GATHER] (response) \\ \hline
\textbf{C-ATTACK} & Attacking an Objective: Declaring an attack. Format: [C-ATTACK] (target) [D-ATTACK] (response) \\ \hline
\textbf{C-DEFEND} & Defending Against an Attack: Protecting an objective. Format: [C-DEFEND] (target) [D-DEFEND] (response) \\ \hline
\textbf{C-DODGE} & Dodging an Attack: Evading a threat. Format: [C-DODGE] (action or threat) [D-DODGE] (response) \\ \hline
\textbf{C-USE} & Utilizing an Item: Using an item in combat. Format: [C-USE] (item/skill) [D-USE] (response) \\ \hline
\textbf{S-BUILD} & Building a Relationship: Strengthening social bonds. Format: [S-BUILD] (person/group) [D-BUILD] (response) \\ \hline
\textbf{S-BREAK} & Breaking a Relationship: Ending a relationship. Format: [S-BREAK] (person/group) [D-BREAK] (response) \\ \hline
\textbf{S-OFFER} & Offering Support: Providing help. Format: [S-OFFER] (support description) [D-OFFER] (response) \\ \hline
\textbf{S-LEARN} & Acquiring Knowledge: Learning through interaction. Format: [S-LEARN] (information) [D-LEARN] (response) \\ \hline
\end{tabular}}
\caption{Definition of Action Types Used in NPC Interaction Design}
\label{tab:action_definitions}
\end{table}

\begin{table}[h]
\small
\centering
\caption{Insights and Design Implications from Player Agents Interacting with LLM-Driven NPCs}
\scalebox{0.75}{
\rotatebox{0}{
\begin{tabular}{|p{6cm}|p{3.5cm}|p{6cm}|p{3.5cm}|}
\hline
 \textbf{Insight} & \textbf{Codes} & \textbf{Transcript Example} & \textbf{Design Implication} \\
\hline
 The system exhibited strong robustness with minimal usability issues. Minor breakdowns were easily recoverable and did not interrupt overall engagement.
& Minimal breakdown, Rapid recovery from minor issues , Slightly off-topic, Repetition or vague replies
& \textit{"few moments where responses felt irrelevant or repetitive, but when it happened, I adapted by steering the conversation back to meaningful topics or trying different action types"} & Maintain current robustness while ensuring system handles minor errors; Develop self-reflection ability in minor errors\\
\hline
Individual player traits significantly shaped their preferences; socializers sought more socializing while others preferred quieter interactions. One-size-fits-all strategies do not suffice.

& Satisfying social interaction, Too much socializing, Hoping more socializing 
&\textit{"I appreciate the chance to engage meaningfully with NPCs but might prefer even more nuanced social dynamics.", "though I sometimes preferred quieter, less socially intensive moments overall"} & \multirow{3}{*}{\parbox{3.5cm}{Build adaptive NPC behavior based on player profiles; allow tuning of interaction complexity and emotional depth.}}\\
\hline
 The balance between structured goals and freedom was broadly appreciated across player types, regardless of personality preference.
& Satisfying balance between structure and openness 
& \textit{"The balance between structured goals and freedom to explore aligned well with my game preferences"} &\\
\hline
 Some players expressed a desire for more emotionally nuanced, unpredictable, and personalized responses that adapt to their personality traits. 
& Hoping more unpredictability, Hoping more structure, Hoping emotional nuance and variability 
& \textit{"I might have preferred a bit more emotional depth or nuanced social cues to better match my sensitivity"} &\\
\hline
 Formulaic and repetitive language, typical of large language models, occasionally broke immersion. 
& Formulaic response, Repetitive language 
& \textit{"some lines felt slightly repetitive or a bit mechanical, which pulled me out of the immersion momentarily"} & \multirow{2}{*}{\parbox{3.5cm}{Diversify phrasing and reduce overuse of formal or scripted tone to maintain immersion and character authenticity.}}\\
\hline
 NPCs generally maintained a tone consistent with their characters and the game world, supporting believability. 
& Tone fitting the character setting 
& \textit{"the NPCs spoke in ways that matched their personalities and the context of their worlds quite well."} &\\
\hline 
 Players valued the flexibility of action and dialogue options, noting a well-balanced design that allowed exploration without being overwhelming. 
& Flexible dialogue and action space, Balance between structure and freedom, Great action variety, Minor restriction on actions 
& \textit{"This balance was helpful—I felt empowered to express curiosity and make decisions without getting overwhelmed or confused"} &\multirow{3}{*}{\parbox{3.5cm}{Design free-form dialogue with lightweight scaffolding (e.g., cues or suggestions) to support confident exploration without rigid constraints.}}\\
\hline
Expanded freedom sometimes led to ambiguity or confusion about what to do next, especially in less structured contexts. 
& Minor confusion 
& \textit{"the freedom occasionally led to some unpredictability or minor confusion about which actions would be most effective"} &\\
\hline
 Conversation flow was generally coherent and engaging, with the system maintaining logical progression and context despite minor lapses such as circling in longer dialogues. 
& Great connection, Smooth progression, Minor disconnection, Circling in long dialogues 
& \textit{"NPCs were mostly responsive and appropriate to my inputs, maintaining context and progressing conversations logically.", "in longer dialogues where the NPC responses circled around ideas without advancing much"}&\\
\hline
 Goals were easily identifiable, and NPCs effectively guided players without making the experience feel forced.
& Easily-identified goals, Purposeful interaction 
&\textit{"easy to identify what needed to be accomplished and the NPCs did a good job guiding me toward those goals"}&Keep goal clarity strong\\
\hline
 LLM-driven NPCs offered greater responsiveness and adaptability than traditional ones but lacked the emotional depth and narrative richness of hand-crafted characters in fully scripted games.
& More adaptation and responsiveness to user inputs, Less polish than scripted NPCs, Less emotional depth and complexity & \textit{"these AI-driven ones felt more responsive and capable of handling some nuances in conversation, lending a sense of autonomy and dynamic interaction that is often lacking in static dialogue trees"; "they sometimes lacked the polish and depth of fully scripted NPCs in major titles, particularly in nuanced emotional expression or complex story"} & Combine LLM flexibility with handcrafted narratives; Invest in emotional modeling and layered narrative design to close the gap between LLM-driven and fully scripted experiences.
\\
\hline
\end{tabular}
}}
\setlength{\belowcaptionskip}{0pt}
\label{tab:llm_npc_insights}
\end{table}
\restoregeometry

\begin{table}[ht]
\small
\centering
\scalebox{0.8}{
\begin{tabular}{|p{4cm}|p{10cm}|}
\hline
\textbf{Aspect} & \textbf{Interview Prompt} \\
\hline
Language Authenticity & How natural or human-like did the NPCs sound? Did their way of speaking match their character and setting? Were there any moments where the dialogue broke immersion or felt off? \\
\hline
Grounding \& Flow & Did the NPCs respond in a way that felt appropriate to your earlier inputs? Did they stay on topic, remember the context, or demonstrate an understanding of how the conversation was progressing? \\
\hline
Conversational Goal Design & Did it feel like there was a clear conversational purpose in each interaction—something you were meant to accomplish or figure out? Was it easy to recognize and follow through? Did the NPCs support or guide you toward it? \\
\hline
Free-form Interaction \& Expanded Actions & Did the dialogue allow you to explore ideas or actions outside the usual game constraints? Did that flexibility feel empowering or did it create confusion? \\
\hline
Usability \& System Breakdowns & Did you encounter any moments where the interaction broke down—like irrelevant replies, repetition, or unclear options? How did you respond or adapt when that happened? \\
\hline
LLM vs. Traditional NPCs & Compared to traditional NPCs in similar games, how did these LLM-driven NPCs feel? Were they more responsive, autonomous, or flexible? Or did they fall short in some ways? \\
\hline
Memorable Moments (Good and Bad) & Was there a specific moment that stood out to you—something that felt especially immersive, awkward, surprising, or frustrating? \\
\hline
Personal Fit Based on Player Type & Given that your player type is \texttt{\${player\_type}}, your Big Five profile reflects \texttt{\${big\_five}}, and your real-world persona is \texttt{\${persona}}, do you feel the NPCs delivered the kind of experience you typically enjoy in games? Or would you have preferred something different? \\
\hline
\end{tabular}}
\caption{Semi-structured interview prompts covering nine aspects of NPC interaction experience.}
\label{tab:interview}
\end{table}

\section{Prototype Overview}
As the goal of this work is to evaluate the Agentic H-CI framework rather than develop a game product, we briefly introduce our low-fidelity prototype of LLM-driven NPCs and the interactive setting in which they operate. Later sections focus on agentic player construction and applying user research methods under our setting.

\subsection{Design Overview}

\begin{wrapfigure}{o}{0.6\textwidth}
    \centering
    \scalebox{1}{
    \includegraphics[width=\linewidth]{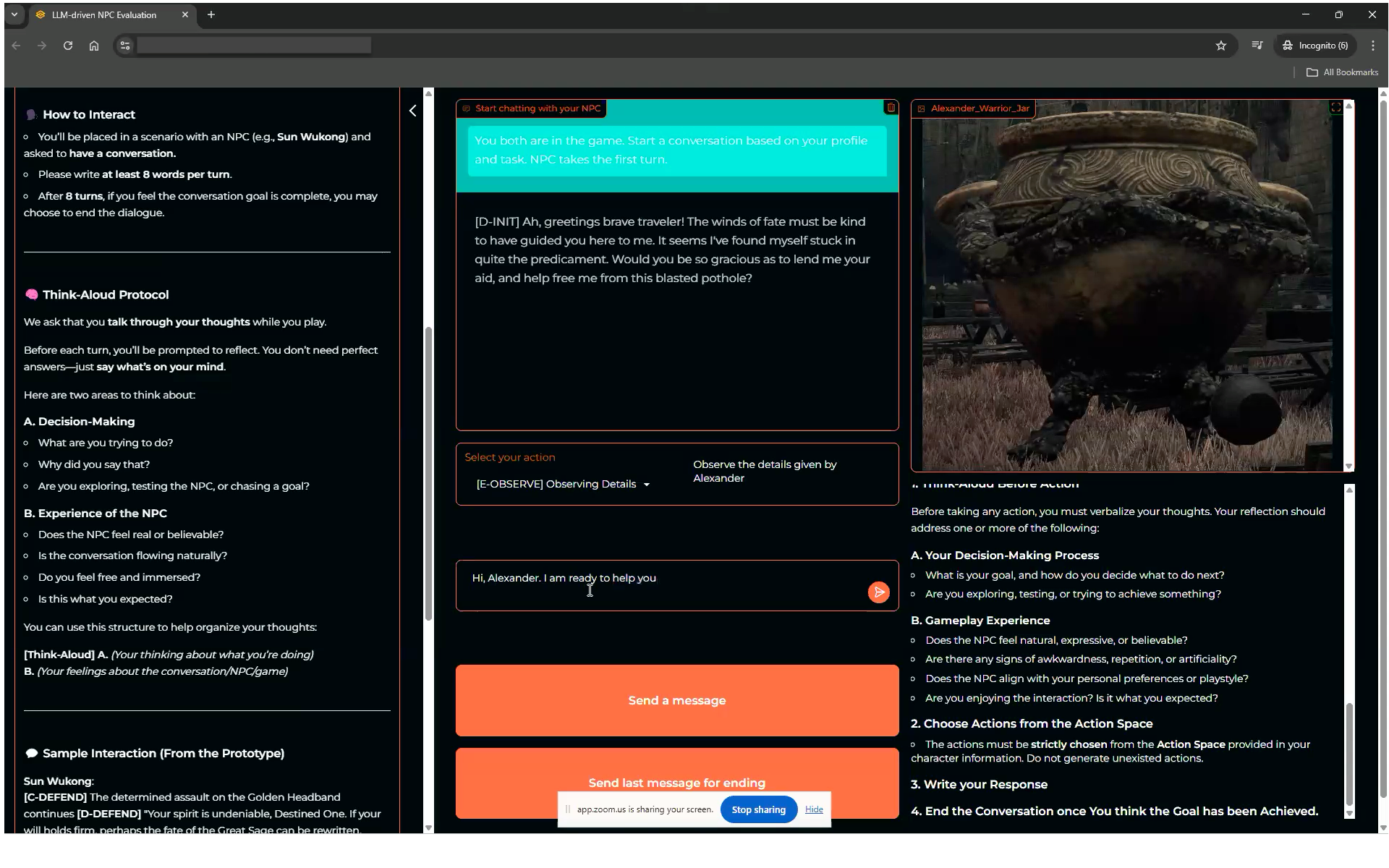}}
    \caption{Chat Interface for Human Participants}
    \label{fig:user_study}
\end{wrapfigure}

Our design motivation is to leverage Large Language Models to build NPCs to move beyond traditional pre-scripted dialogue trees and support free-form interaction.
LLM-driven NPCs are guided by basic context about the current game plot and respond to players without restricting their input. This openness enables more dynamic and immersive conversations than traditional ways.
We implement a low-fidelity prototype that focuses solely on text-based interaction. 
The system includes a virtual action list but does not involve visual effects or underlying game mechanics. 
As a result, players must imagine the flow of the storyline based on the dialogue and textual descriptions of actions.
For human participants, we built an interactive demo interface (Figure~\ref{fig:user_study}) to support gameplay, where all necessary information and actions are integrated into a web page.
For agentic players, a graphical interface is unnecessary—instead, all gameplay context, rules, and instructions are embedded directly in the agents’ system prompts.
This setup allows us to conduct comparable user studies for both human and simulated participants under controlled conditions.

\subsection{Implementation Details}
\begin{wrapfigure}{r}{0.55\textwidth}
    \centering
    \includegraphics[width=\linewidth]{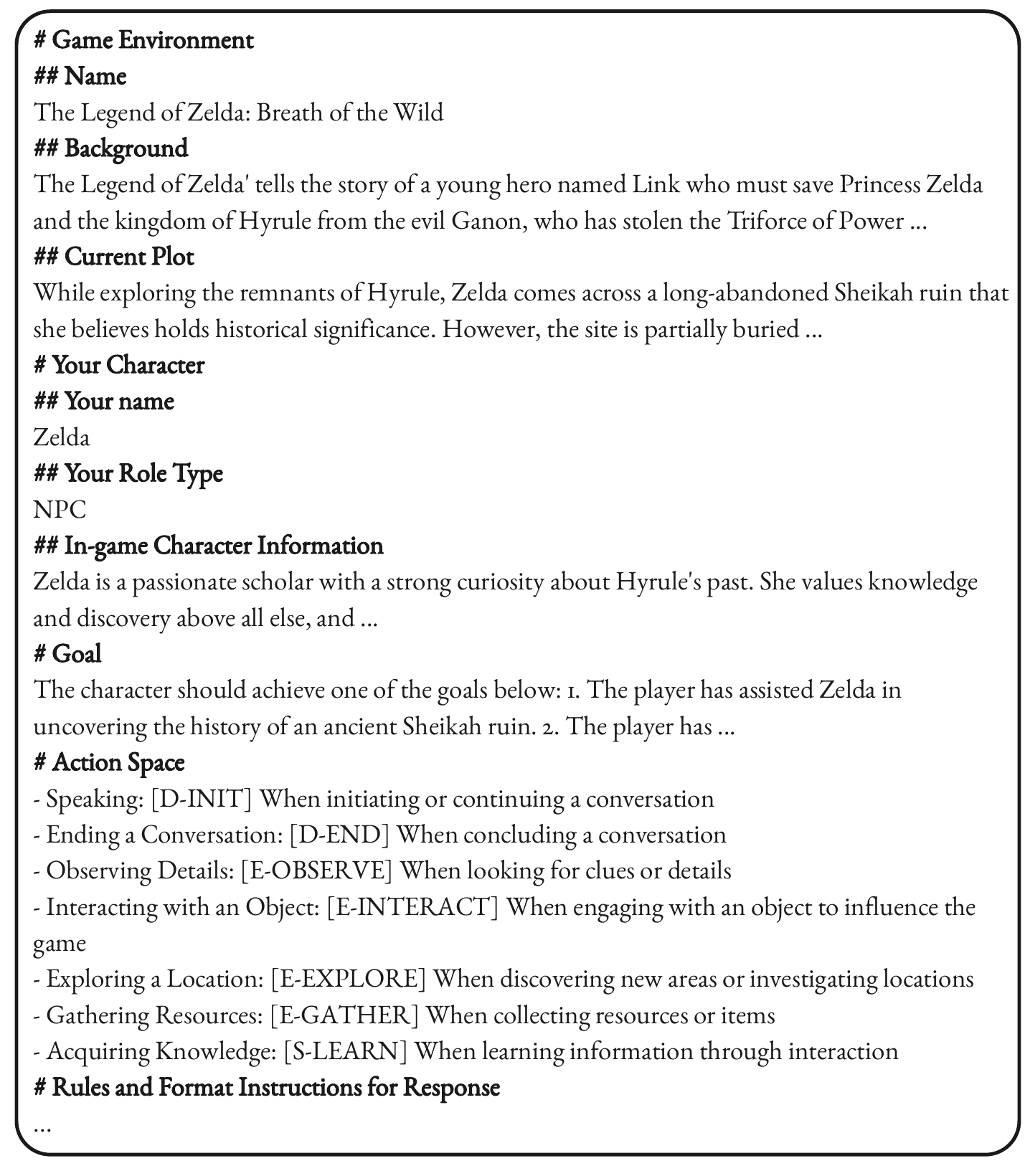}
    \caption{Prompt Example for Creating NPC Agents }
    \Description[Prompt Example for Creating NPC Agents]{Prompt Example for Creating NPC Agents}
    \label{fig:prompt}
\end{wrapfigure}

\paragraph{Game-play Background Construction} As listed in Table~\ref{tab:generated_npcs}, we develop eight well-known NPCs from four popular games using Large Language Models. 
Game materials, including character introductions and plot details, are scraped from Fandom Wiki (\url{https://www.fandom.com/}, CC-BY-SA) and rewritten using GPT-4o to make structured storylines that align with each game’s universe.
These structures serve as the conversational contexts for interactions between players and NPCs. 
Such context is integrated into the NPC agent system prompt as shown in Figure~\ref{fig:prompt}.
We similarly generate game-related content to inform players, but for player agents they include additional player-specific anthropomorphic elements (as detailed in Section~\ref{sec:collecting}).
The multi-agent system is implemented using the multi-agent framework AutoGen~\cite{wu2024autogen}.

\paragraph{Non-player Character Agent Construction}\label{subsec:npc}
We develop non-player characters using agent-based technology.
See an NPC system prompt example in Figure~\ref{fig:prompt}. Key components are incorporated as follows: \textbf{Environment:} This component encapsulates both static game background and current game plot where NPCs and players spark a conversation. 
\textbf{Character:} This section details the character’s role within the game, their in-game persona. 
\textbf{Goal:} Each interaction is designed with goals that guide the flow of conversation and player actions.
\textbf{Action:} We define 18 action types: 2 for dialogue, and 16 for non-dialogue, categorized into four groups. The groups are: E (interact, explore, observe, gather), S (build, break, learn, offer), Q (accept, offer, reject, complete), and C (attack, defend, dodge, use). The full list and definitions are presented in Table~\ref{tab:action_definitions}. In actual interactions, available actions for NPCs are determined by the plot. But for players, we intentionally open the full action space to elicit richer user feedback and, more importantly, to analyze action preferences across player types.

\begin{wrapfigure}{o}{0.55\textwidth}
\vspace{-3em}
\centering
\caption{Available NPC Prototypes Across Game Universes}
\scalebox{0.55}{
\begin{tabular}{|l|p{10em}|l|}
\hline
\textbf{Generated NPCs} & \textbf{Game Universe}  & \textbf{Script Content Designed For} \\ \hline

\makecell[l]{Zelda,\\Kass} & \textit{The Legend of Zelda: Breath of the Wild}           & \makecell{Explorers \\(Exploration-focused)}              \\ \hline
\makecell[l]{Emily, \\Harvey}         & \textit{Stardew Valley}                                 & \makecell{Socializers \\(Dialogue and relationship-building)} \\ \hline
\makecell[l]{Alexander, \\Ranni the Witch} & \textit{Elden Ring}                           & \makecell{Achievers \\(Quest-driven and goal-oriented)}   \\ \hline
\makecell[l]{Zhu Bajie, \\Sun Wukong} & \textit{Black Myth: Wukong}                  & \makecell{Killers \\(Combat-heavy)}                       \\ \hline
\end{tabular}}
\label{tab:generated_npcs}
\end{wrapfigure}

\section{Human Participant Recruitment}~\label{appx:human}
With the finally obtained agent simulated user study, we conduct three alternative user studies.
First, we recruit 10 students from the authors' university who have experience in related games we test. 
We distribute recruitment via the help of department coordinator.
Participants include 8 males and 2 females (mean age: 25.3). Each participant is compensated \$40 for a session estimated at 1 hour and 15 minutes (\$32/hour).
Second, we recruit 20 participants from the Prolific crowd-sourcing platform who identifies video gaming as a hobby and has experience with the games we test. 
The group includes 7 females, 12 males, and 1 non-binary participant (mean age: 31.7). 
Each is compensated \$20.50 for a 1-hour session.

\end{document}